# Optical Lattices with Higher-order Exceptional Points by Non-Hermitian Coupling

Xingping Zhou,[1][2] Samit Kumar Gupta,[1][3] Zhong Huang,[1][2] Zhendong Yan,[1][2] Peng Zhan,[1][2][4]* Zhuo Chen,[1][2][4] Minghui Lu,[1][3][4],* and Zhenlin Wang[1][2][4]*,

[1] *National Laboratory of Solid State Microstructures, Nanjing University, Nanjing 210093, China*

[2] *School of Physics, Nanjing University, Nanjing 210093, China*

[3] *College of Engineering and Applied Sciences, Nanjing University, Nanjing 210093, China*

[4] *Collaborative Innovation Center of Advanced Microstructures, Nanjing University, Nanjing 210093, China*

*\* zhanpeng@nju.edu.cn*

*\* luminghui@nju.edu.cn*

*\*zlwang@nju.edu.cn*

Exceptional points (EPs) are degeneracies in open wave systems with coalescence of at least two energy levels and their corresponding eigenstates. In higher dimensions, more complex EP physics not found in two-state systems is observed. We consider the emergence and interaction of multiple EPs in a four coupled optical waveguides system by non-Hermitian coupling showing a unique EP formation pattern in a phase diagram. In addition, absolute phase rigidities are computed to show the mixing of the different states in definite parameter regimes. Our results could be potentially important for developing further understanding of EP physics in higher dimensions via generalized paradigm of non-Hermitian coupling for a new generation of parity-time (*PT*) devices.

It is well-known that systems with open boundaries [1] or material loss and gain [2] can be described by Hamiltonians that are non-Hermitian. Non-Hermitian systems have recently attracted great scientific interests, both theoretically and experimentally, in open systems with energy gain or loss. A non-Hermitian Hamiltonian can exhibit many of intriguing phenomena beyond that of a Hermitian system. Consequently, researchers have become increasingly aware of the potential effects and applications reminiscent of the non-Hermitian systems. For instance, a non-Hermitian Hamiltonian have non-orthogonal eigenfunctions with complex eigenvalues where the imaginary part corresponding to decay or growth [2]. At certain points in parameter space fin as exceptional points (EPs) [3] or non-Hermitian singularities [4, 5], two (or more) eigenfunctions collapse into one, so the eigenspace no longer forms a complete basis, and these eigenfunctions become self-orthogonal under the unconjugated *inner product*. The concept of EP also features prominently in the study of parity-time (*PT*) symmetric optics[6, 7], and many interesting features have been discovered theoretically and demonstrated experimentally, such as topological characteristics [8-10], loss-induced transparency [11], coherent perfect absorption [12], slow light [13], non-Hermiticity induced flatbands [14-16] and Bloch oscillations [17-19]. *PT* structures are composed of judiciously distributed gain and loss mechanisms, and the loss is usually generated by metal coating or open boundary for radiation, while the gain is created using optical or electrical pumping. They feature a form of symmetry breaking between '*PT*-symmetric' phases (with real eigenvalues) and '*PT*-broken' phases (with conjugate pair eigenvalues), separated by EPs where two (or more) eigenstates become degenerate. Very recently, the emergence and interaction of higher-order EPs in a four-state system using coupled acoustic cavities have been reported by K. Ding et al., which presents more interesting topological characteristics and richer EP physics [8].

On the other hand, coupled optical waveguides provide a platform for exploring the physics of EPs and *PT* symmetry, because they can be fabricated with precise control over the structural parameters affecting propagation constants and coupling coefficient. Klaiman et al. proposed a setup of two coupled *PT*-symmetric waveguides with complex refractive index for the visualization of second-order EPs [2]. The predictions received convincing experimental confirmations in 2009 by Guo et al. [11] and in 2010 by Rüter et al. [20]. The imaginary parts are interpreted as gain (loss) of the field intensity, e.g., by optical pumping and absorption and their strength control the degree of non-Hermiticity. For multiple waveguides systems, the interesting physics arising from non-Hermiticity is not limited to those EPs that are induced by a two-waveguide system. Recently, the dark state [14, 21] and the *PT* symmetry recovery behaviors [22, 23] have been found in three-waveguide and four-waveguide system, respectively. Furthermore, a *PT*-symmetric topological interface state has been experimentally demonstrated in a non-Hermitian waveguide arrays system [24].

Most studies of EP or higher-order EPs [25] and their intriguing physical effects associated with *PT*-symmetry rely on on-site material gain/loss or open boundaries. In an attempt to develop a generalized description of complex EP physics, non-Hermitian couplings between the coupled waveguides, which was proposed by Leykam et al., describe a situation where the mode amplitude undergoes gain or loss while hopping between the sites [26]. Non-Hermitian coupling can also be regarded as the phase of a hopping amplitude [27]. It is worth mentioning that non-Hermitian couplings can be realized by a single open channel consisting of two optical dipole antennas [28], embedding amplifying or lossy media between adjacent waveguides [26, 29, 30], periodically modulating the onsite gain/loss [27], or as an effective description of larger waveguide networks [31]. In addition, non-Hermitian coupling can appear naturally in the Bloch Hamiltonian of a periodic non-Hermitian system [15, 32]. In fact, adding more non-Hermitian components to a Hamiltonian results in further changes in its mathematical form and more degrees of freedom. So systems with non-Hermitian couplings give rise to physical effects such as the fractional value of winding number [33] and the flatband [26]. In this paper, we present the emergence and coalescence of higher-order EPs and power oscillations in a *PT* optical four-waveguide system that involves non-Hermitian couplings. The emergence and coalescence of higher-order EPs can be summarized in a phase diagram featuring an exceptional point formation pattern (EPFP). The coalescence of higher-order EPs forms two curves in the parameter space, partitioning the phase space into three regions each with a unique EPFP. Our theoretical analysis is based on coupled-mode theory (CMT).

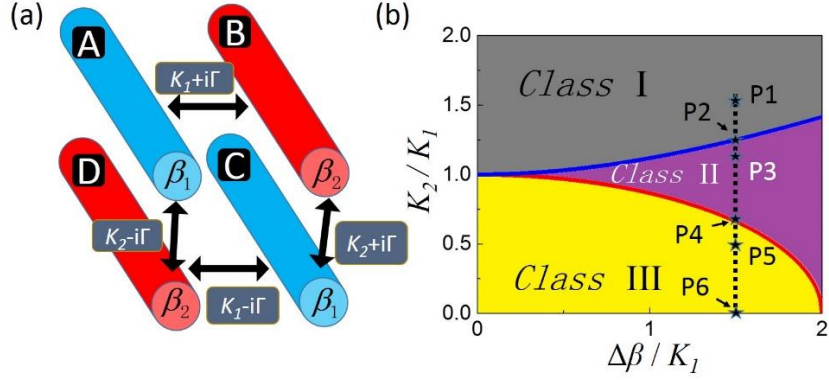

Fig. 1. (a) Schematic of the setup, consisting of four waveguides (labeled A - D) with non-Hermitian couplings. (b) Phase diagram in the $K_2 \sim \Delta\beta$ space with $\Delta\beta = \beta_2 - \beta_1$. The gray, purple and yellow regions represent the distinct EPFP marked with class I, II, and III, respectively. The solid blue and red lines mark the coalescence of higher-order EPs. The vertical black dashed line and the six points (labeled from P1 to P6) are drawn to show the EPFPs in Fig. 2.

We construct a four-state system as shown in Fig. 1(a), which consists of four waveguides (labeled as A, B, C and D). Waveguides A and C (colored blue) form one pair with propagation constant $\beta_1$, and waveguides B and D (colored red) form another pair with propagation constant $\beta_2$. The Hamiltonian of the system can be written as

$$H = \begin{pmatrix} \beta_1 & K_1+i\Gamma & 0 & K_2-i\Gamma \\ K_1+i\Gamma & \beta_2 & K_2+i\Gamma & 0 \\ 0 & K_2+i\Gamma & \beta_1 & K_1-i\Gamma \\ K_2-i\Gamma & 0 & K_1-i\Gamma & \beta_2 \end{pmatrix} \quad (1)$$

The diagonal terms of $H$ describe the propagation constants of the sites (waveguides). For simplicity, the $\beta_1, \beta_2, K_1, K_2,$ and $\Gamma$ are set to real numbers here. The off-diagonal terms $\kappa_{nm}$ describe the coupling between the $n$th and $m$th sites. The non-Hermitian couplings are $\kappa_{12} = \kappa_{21} = K_1+i\Gamma = \kappa_{34}{}^* = \kappa_{43}{}^*$ and $\kappa_{32} = \kappa_{23} = K_2+i\Gamma = \kappa_{14}{}^* = \kappa_{41}{}^*$. Each coupling term consists of a Hermitian part, $K_1$ or $K_2$, and a non-Hermitian part, $\Gamma$. The eigenvalues of the Eq. (1) take the following form:

$$\beta_{eig} = \beta_0 \pm \sqrt{\Delta_1 \pm 2\sqrt{\Delta_2}} \quad (2)$$

where $\beta_0 = (\beta_1+\beta_2)/2$ and

$$\Delta_1 = K_1^2 + K_2^2 - 2\Gamma^2 + \frac{1}{4}\Delta\beta^2 \tag{3}$$

$$\Delta_2 = (-K_1^2 + \Gamma^2)(-K_2^2 + \Gamma^2) \tag{4}$$

with $\Delta\beta = \beta_1 - \beta_2$.

From Eq. (2), it is obvious that coalescence of states (CSs) [8] could occur under three conditions depending on the parameter values of $K_1$, $K_2$, $\Delta\beta$ and $\Gamma$, which we refer to as a single EP2: $\Delta_1 \pm 2\sqrt{\Delta_2} = 0$, $\Delta_1 \neq 0$, $\Delta_2 \neq 0$, two EP2's: $\Delta_1 \neq 0$, $\Delta_2 = 0$ and EP4: $\Delta_1 = \Delta_2 = 0$, respectively. A single EP2 corresponds to a normal EP with one state defective and two EP2's corresponds to two EPs with two states defectives. At EP4, four states coalesce with three states defective.

When $\Gamma / K_1$ increases, $\Delta_1 \pm 2\sqrt{\Delta_2}$ and $\Delta_2$ becomes negative at particular points, leading inexorably to a broken phase with complex conjugate eigenvalues. Then solving the EP4 singularity under the condition of $\Delta_1 = \Delta_2 = 0$ gives the following equations:

$$K_2 = \sqrt{-\frac{1}{4}\Delta\beta^2 + K_1^2} \tag{5}$$

$$K_2 = \sqrt{\frac{1}{4}\Delta\beta^2 + K_1^2} \tag{6}$$

Equation (5) and equation (6) are plotted by a solid red line and a solid blue line in Fig. 1(b), respectively. From Eq. (3) and (4), we see that, depending on the parameters $K_1$, $K_2$, and $\Delta\beta$, different combinations of a single EP2, two EP2's, and EP4 may appear in the EPFP when $\Gamma$ increases continuously. Figure 1(b) shows a phase diagram in the space of two dimensionless parameters $K_2 / K_1$ and $\Delta\beta / K_1$. Three regions exist, termed as class I, II, and III, with their boundaries marked by the solid red and blue line. Each class represents a distinct EPFP, in which the EPs have different singularity types. The four coupled optical waveguides system by non-Hermitian coupling with multiple EPs considered here can be used for mode selection, so it is worthwhile to note that the optical waveguides system with multiple EPs considered here might be used as an effective tool for modal demultiplexing [34].

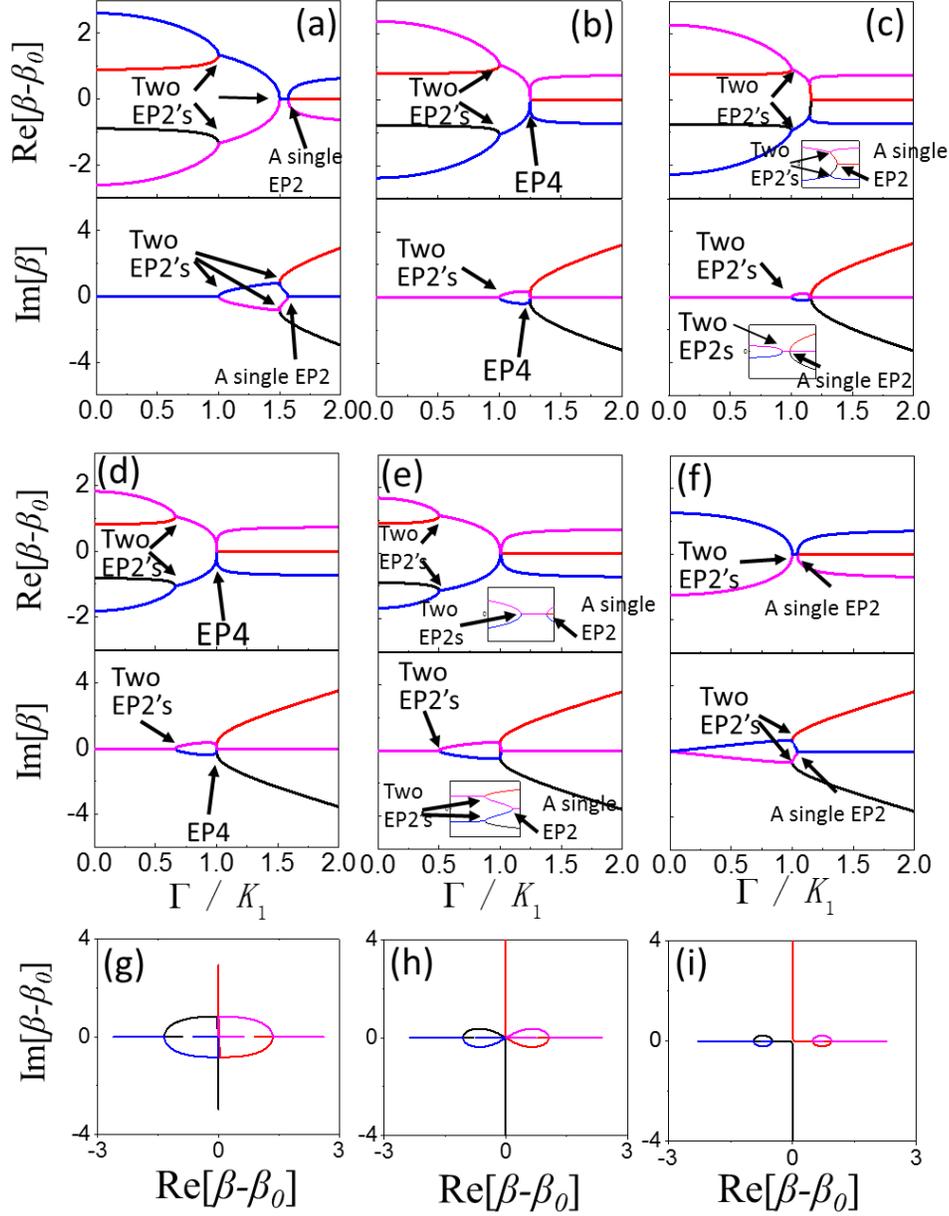

Fig. 2. Real and imaginary parts of eigenvalues as function of $\Gamma/K_1$ for points P1 to P6 in the phase diagram shown in Fig. 1(b) are plotted from (a) to (f). Each EP is labeled to denote its type as evident in the main text of the article. The insets are zoom-in view of the eigenvalues. Trajectories of H eigenvalues in the complex plane with increasing non-Hermitian parameter $\Gamma$ depicted in (g)-(i) that correspond to Fig. 2 (a)-(c).

We choose a value of $\Delta\beta/K_1 = 1.5$ and then decrease $K_2/K_1$ continuously from 1.5 (point P1) to 0 (point P6), as marked by the dotted black line in Fig. 1(b). The point P1 lies in class I, which occupies the top area of the diagram. We can see from Fig. 2(a) that increasing $\Gamma/K_1$ renders $\Delta_2$ vanish and consequently we have two EP2's singularities in the spectrum with complex eigenvalues. Further increasing of $\Gamma/K_1$ yields another root of $\Delta_2$ which forms two EP2's singularity. Upon a further increase in $\Gamma/K_1$, $\Delta_1 - 2\sqrt{\Delta_2}$ becomes negative and a single EP2 singularity appears. Decreasing the $K_2/K_1$ will bring the system to the blue solid line in Fig. 1(b), which is given by Eq. (6). As shown in Fig. 2(b) (point P2), the configurations on this blue line always carry a EP4 type singularity for some particular values of $\Gamma/K_1$ and three states are defective at EP4 (same as Fig. 2(d)). Such a singularity is a higher-order EP. The blue line hence represents a line of higher-order singularities. In addition to the EP4 singularity, we find another two EP2's singularities at a smaller value of $\Gamma/K_1$ around 1. Below the blue line in Fig. 1(b) (in the class II), the EP4 singularity splits into three singularities: two EP2's on the left and a single EP2 on the right. A typical EPFP of class II is shown in Fig. 2(c) (point P3). Interestingly, we find symmetry recovery [20] between the two EP2's singularities and a single EP2 singularity indicated in the inset of Fig. 2(c). In this area, all eigenvalues are real again with $\Gamma/K_1$ increasing and the symmetry is recovered. The red solid line in Fig. 1(b), which is given by Eq. (5) separates class II from class III. In similar to the point P2, point P4 on the red line carries an EP4 type singularity. Below this line (in the class III), the EP4 singularity also splits into three singularities. In this case shown in Fig. 2(e), when the symmetry is broken at the two EP2's with a smaller value of $\Gamma/K_1$, the eigenvalues do not become real again and symmetry recovery is no longer possible. At $K_2/K_1 = 0$ (point P6), we find that the Hamiltonian is non-Hermitian and the entire spectrum is imaginary when $\Gamma/K_1 \neq 0$. There are only two EP2's singularities on the left and a single EP2 singularity on the right when $\Gamma/K_1 > 0$ in this case. In addition, the most interesting thing is that in our system EP4 can split into three EP2's either on the on the imaginary axis (Fig. 2(g)) or on the real axis (Fig. 2(i)) while all the other cases can be categorized by how this EP4 is split in the complex plane. This phenomenon is different from the splitting of an EP3 into two EP2's discussed in Ref [35]. It is worthwhile to mention that strictly speaking our system possesses $RT$-symmetry instead of $PT$-symmetry [23, 36] satisfying $[H, RT] = 0$, where $R$ denotes rotation by $\pi$ about the central axis of the system. The phase transition behaviors of $RT$ systems could be very distinct as compared to that of $PT$. For instance, one can have a situation where the spectrum for $RT$ type of system is completely real in contrast to the corresponding PT variant where it is not so. Besides $RT$ symmetry, our system has non-Hermitian particle-hole (NHPH) symmetry[37]. The effective

Hamiltonian of our system satisfies $\{H-\beta_0 1, CT\} = 0$. The eigenstates can exist in two phases i.e., the symmetric phase where $\beta_n = e^{i\theta_n} CT\beta_n, \beta_n = -\beta_n^*$ and the broken-symmetry phase where $\beta_m = CT\beta_n, \beta_m = -\beta_n^* (m \neq n)$.

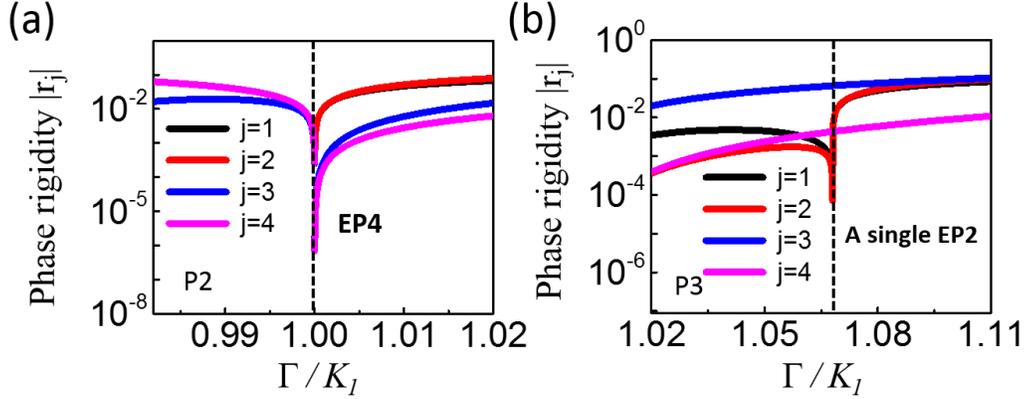

Fig. 3. Phase rigidity of all the eigenstates as functions of $\Gamma / K_1$ for points P2 and P3 in the phase diagram shown in Fig. 2.

To identify the singularity order, we plot in Fig. 3 the absolute value of phase rigidity [8], defined as

$$r_j = \frac{<\phi_j^L | \phi_j^R>}{<\phi_j^R | \phi_j^R>} \tag{7}$$

for each state $j$ as a function of $\Gamma / K_1$ for the point P2 and P3 in the phase diagram, where $\phi^R$ and $\phi^L$ are right and left eigenvectors of Hamiltonian. Phase rigidity is a measure of the mixing of different states. In the absence of $\Gamma / K_1$, all four states are distinct and their phase rigidity is close to unity. As $\Gamma / K_1$ is increased, phase rigidities are reduced as some states begin to mix. In Fig. 3(a), we plot the absolute value of phase rigidity of point P2 in the phase diagram. In EP4 ($\Gamma / K_1 \approx 1$), all four states have zero rigidity, which indicates they are all linearly dependent with three defective states. Two states are completely mixed at the EP where the phase rigidity vanishes as shown in Fig. 3(b). It is clear that $|r_j|$ vanishes for states $j = 1$ and 2 at the linear crossing point at $\Gamma / K_1 \approx 1.06$, indicating a defective state.

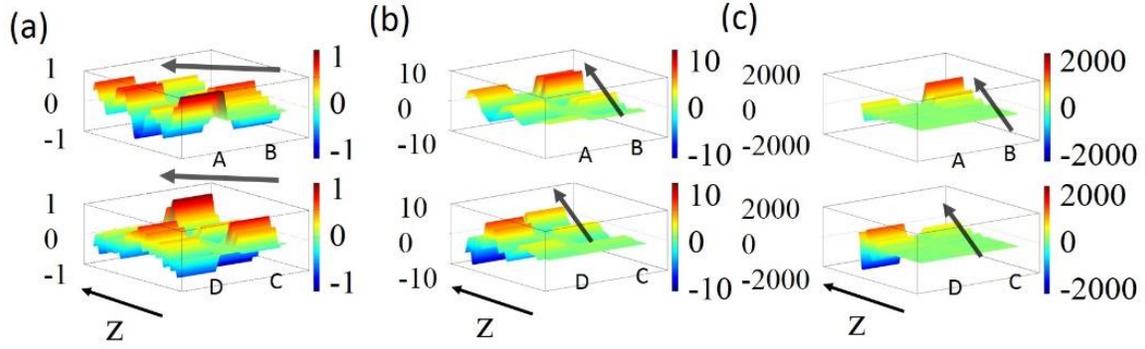

Fig. 4. The amplitudes distributions of the four-waveguide system for the propagating total fields based on coupled mode theory. The lower panels refer to the density plots of the field distributions corresponding to the upper ones.

In order to confirm the underlying physics of higher-order EPs in the four-waveguide system, we perform theoretical calculations based on CMT. The field propagation within the waveguides can be described by the following set of equations in the tight-binding approximation (TBA):

$$\begin{aligned}
-i\frac{\partial}{\partial z}\psi_A &= \beta_1\psi_A + (K_1+i\Gamma)\psi_B + (K_2-i\Gamma)\psi_D \\
-i\frac{\partial}{\partial z}\psi_B &= \beta_2\psi_B + (K_1+i\Gamma)\psi_A + (K_2+i\Gamma)\psi_C \\
-i\frac{\partial}{\partial z}\psi_C &= \beta_1\psi_C + (K_2+i\Gamma)\psi_B + (K_1-i\Gamma)\psi_D \\
-i\frac{\partial}{\partial z}\psi_D &= \beta_2\psi_D + (K_1-i\Gamma)\psi_C + (K_2-i\Gamma)\psi_A
\end{aligned} \qquad (8)$$

where $\Psi_A$, $\Psi_B$, $\Psi_C$, and $\Psi_D$, are the amplitudes of the light field in these four waveguides, respectively. The initial incident amplitude is $\Psi(0) = (1, 0, 0, 0)$, which means the waveguide labeled A being excited at the input of the array. Fig. 4 shows the calculation results of three cases. For the sake of simplicity, we set $\beta_1 = 1$ and $K_1 = 1$. We choose the parameters corresponding to the point P2, for which the real and imaginary parts of the eigenvalues are presented in Fig. 2(b). Firstly, we set $\Gamma / K_1 = 0.5$ which is *PT*-symmetric with real eigenvalues (propagation constants) and has eigenstates (waveguide supermodes) that respect *PT* symmetry. The real spectrum in the *PT*-symmetric phase indicates that wave propagation with conserved energy can be realized in the presence of non-Hermitian coupling as shown in Fig. 4(a). As the value of $\Gamma / K_1$ approaches the EPs, the system become *PT*-broken characterized by complex conjugate propagation constants, together with supermodes that do not respect *PT* symmetry. At the EP and higher-order EPs, the total beam power starts to grow exponentially (see the corresponding color bars) as presented in Fig. 4(b) (the EP with $\Gamma / K_1 = 1$) and Fig. 4(c) (the higher-order EP with $\Gamma / K_1 = 1.25$).

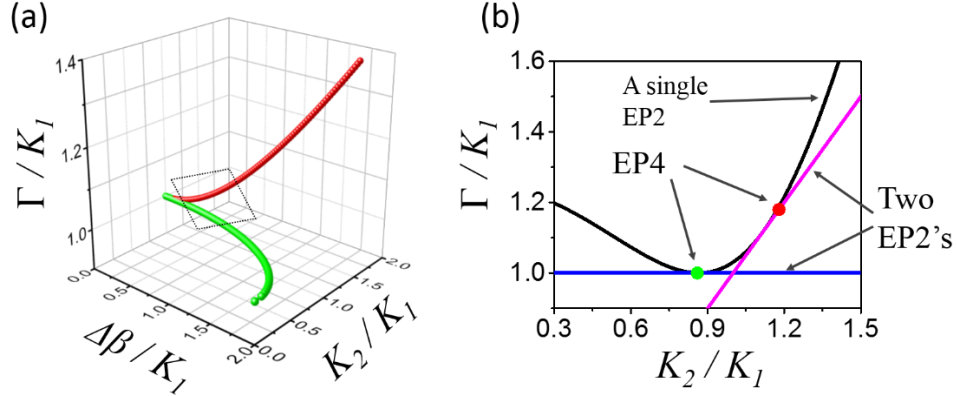

Fig. 5. (a) Curves denoting the positions in three-dimensional parameter space ($\Delta\beta$, $K_2$, $\Gamma$) where EP4 are found for the waveguides system in Fig. 1(a). (b) Projections of curves in the two-dimensional parameter space ($K_2$, $\Gamma$) of the dotted rectangle in Fig. 4(a).

Higher-order EPs are related to the parameter sensitivity of eigenmodes. Despite the fact that it has so far attracted little attention, it is worthwhile to mention that it could be imperative in an expected experimental confirmation. To find out curves of higher-order EPs in the system, we use the condition of $\Delta_1 = \Delta_2 = 0$. We illustrate the results in Fig. 5(a) for a configuration of the system. The points on the two space curves (colored green and red, respectively) satisfy the condition of higher-order EPs. To find the curves of lower-order EPs in the system, we select a plane (indicated by the dotted rectangle) of $\Delta\beta / K_1 = 1$ in the three-dimensional parameter space. Solving the conditions of $\Delta_1 \pm 2\sqrt{\Delta_2} = 0$ and $\Delta\beta / K_1 = 1$, we get the solid black line in Fig. 5(b) which indicates a single EP2 singularity. The solid pink and blue lines are given by $\Delta_2 = 0$, and $\Delta\beta / K_1 = 1$, respectively, which indicates the two EP2's singularity. The two points (green and red dots) that refer to the points of intersection between the black and blue lines, and the black and pink lines are the EP4 singularities. So the higher-order EPs can be seen as result of a coalescence of lower-order EPs.

We have studied a closed-form four-waveguide system with non-Hermitian coupling which can serve as a promising setup for verification of higher-order EPs. The emergence and coalescence of higher-order EPs induced by non-Hermitian coupling have been studied numerically. We have shown the mixing of the states by computing absolute phase rigidities. Beam power propagations in the non-Hermitian system of four-waveguides with non-Hermitian couplings are presented by CMT method. We also demonstrate the curves of the lower-order and higher-order EPs in the parameter space of the system. Our work present a generalized description of higher-order EPs though non-Hermitian couplings in coupled optical waveguides system which could be potentially important in various aspects of practical optical settings, such as modal demultiplexing, sensing and so on. In non-Hermitian system, the eigenfrequency splitting $\Delta\omega$ can be accentuated by orders of magnitude, because it follows an $\varepsilon^{1/N}$-dependence [25] (N represents the order of the exceptional point, $\varepsilon$ is the small disturbance). The numerical computation of fourth-order EP in our system could be suggestive of enhanced sensitivity of such few-site optical waveguides system upon judicious device engineering and tuning. In addition, the presence of multiple EPs in this system can be useful for enhanced prospect of mode selectivity and modal demultiplexing.

**Acknowledgements**

The authors thank for the support by National Key R&D Program of China (2017YFA0303702); National Science Foundation of China under Grant Nos. 11674166, 11674168, 11774162 and 11621091.